\documentclass[amsmat,amssymb,amsfonts,aps,prb,twocolumn,showpacs]{revtex4}

\usepackage{graphicx}
\usepackage{dcolumn}
\usepackage{bm}
\begin{document}

\title{Theory of single spin  inelastic tunneling spectroscopy.}

\author{J. Fern\'andez-Rossier}
\affiliation{Departamento de F\'{\i}sica Aplicada,
Universidad de Alicante, San Vicente del Raspeig, 03690 Spain }

\date{\today}

\begin{abstract}

Recent work  shows that  inelastic electron scanning tunneling microscope (STM)      probes the elementary spin excitations of  a single and  a few  magnetic atoms in  a thin  insulating layer. Here I show that  these new type of spectroscopy is
described using   a phenomenological spin-assisted tunneling Hamiltonian.
Within this formalism, the  inelastic $dI/dV$ lineshape is related to  the  spin spectral weight of the probed magnetic atom. This accounts for the spin selection rules observed experimentally. The theory agrees well 
with existing STM experiments for single Fe and Mn atoms as well as  linear chains a few   Mn atoms. The  magnetic anisotropy in the inelastic $dI/dV$ 
and the    marked odd-even $N$ effects are accounted for by the theory.

\end{abstract}

 \maketitle

Electron tunneling is  one of the central themes in  condensed matter physics.
It lies behind fundamental phenomena like, Josephson effect\cite{Josephson} and tunneling magnetoresistance\cite{TMR},
 and   provides an extremely versatile spectroscopic tool, 
both in tunneling junction \cite{Jaklevic66,Tedrow}  and   STM  geometries \cite{STM}.
The use of inelastic electron tunneling  to determine  the 
vibration spectra of ensembles of molecules inside tunnel barriers goes back
to the seminal work of Jaklevik and Lambe\cite{Jaklevic66}. They observed steps in the 
 differential conductance $dI/dV$ curve at particular values of the bias voltages which matched 
the vibrational energy spectra of different molecules. This led to the notion
of inelastic assisted tunneling\cite{Jaklevic66,Scalapino}: an electron  could tunnel across the barrier giving away 
its excess  energy $eV$ to create an elementary excitation.  In this framework,  
as  $eV$ increases, new inelastic transport channels open, resulting in steps in the $dI/dV$ curve.

With the advent of the STM, it has been possible to downscale the technique of   inelastic tunneling vibrational spectroscopy to the single molecule level \cite{Ho}, a possibility anticipated in the early days of STM \cite{BGR}.
In a series of striking experiments\cite{Heinrich1,Heinrich2,Heinrich3}
 Heinrich {\em et al} have used  inelastic STM spectroscopy to probe 
 the spin flip excitations of a single and a few  transition metal  atoms in a surface by means of STM spectroscopy. They have measured the single Mn atom Zeeman gap \cite{Heinrich1,Heinrich3}, the collective spin  excitations of chains of up to 10 Mn atoms \cite{Heinrich2} and the spin flip transitions within the ground state manifold of a single iron atom, split due to the single atom  magnetic anisotropy\cite{Heinrich3}. 
Analogously,  Xue {\em et al.} to have used inelastic STM spectroscopy to probe
the spin excitations of one and a few Cobalt Phthalocyanines and have measured their
exchange coupling\cite{Xue08}.

\begin{figure}
[hbt]
\includegraphics[width=3.2in]{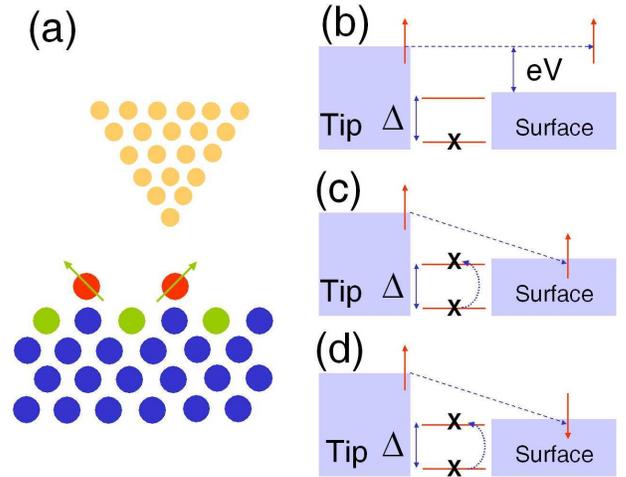}
\caption{ \label{fig1}(Color online). (a) Scheme of the experimental setup
(b) Elastic tunneling process. (c) Inelastic spin assisted tunneling process. (d) Inelastic spin flip tunneling process. }
\end{figure}

Therefore,  spin  assisted inelastic tunneling spectroscopy (SITS) provides a direct measurement of the spin dynamics of one or a few magnetic atoms, and complements 
spin polarised scanning tunneling spectroscopy \cite{Wiesendanger} which is sensitive to the average relative orientation of the  magnetic moments of tip and surface ($\vec{m}_T$ and $\vec{m}_S$ respectively). 
SITS   permits to measure  energy scales  like interatomic  exchange $J$ \cite{Heinrich2,Xue08}, g factor\cite{Heinrich1,Heinrich3,Xue08}  and magnetic anisotropy tensor \cite{Heinrich3}. These quantities 
determine the spin dynamics of the  magnetic atoms(s).  
In contrast to the case of vibrational spectroscopy\cite{Scalapino}, the physical origin of   coupling between the transport electrons and the local spins which makes SITS possible is not clear\cite{persson}. Hirjibehendin {\em et al}  mention two possibilities\cite{Heinrich3},  exchange or dipolar coupling, although the former encompasses   a variety of different mechanisms, like  direct, kinetic, etc.
The coupling must account for a number of experimental observations. 
The height of the steps in the $dI/dV$  scales like the sum of the squares of the matrix elements between the initial and final states of the operators $S_{a}$ with $a=x,y,z$,  the spin  of the atom probed by the SITS\cite{Heinrich3}. This is related to the selection rule for the change of the spin of the local spin $\delta S_z=\pm 1, 0$   \cite{Heinrich2,Xue08}. 
Therefore, there is a relation between  the inelastic current and the {\em spin spectral weight} ${\cal S}_{aa}(\omega)$ of the magnetic atom(s). 
In this paper I  show that an effective spin-assisted tunneling  Hamiltonian 
 naturally explains the relation between the inelastic current and the  spin spectral weight   ${\cal S}_{aa}(\omega)$  and accounts for the main experimental findings.

The experimental system consists of an insulating thin layer deposited  on a metallic surface (see fig. 1a). Magnetic atoms lie in the insulating layer and are probed by a STM . A natural model for this system would thus feature   3 types of fermion operators, tip, surface and insulating layer, plus  the spin operator of the magnetic atoms.  In such an approach, the current, evaluated to lowest order in the tunneling coupling, is related to to the spectral function of the transport electron in the insulating layer interacting with the local spins \cite{JFR07}. This is different from the experimental findings described above. In particular, the conductance so evaluated would have a Coulomb Blockade gap unless the central region is in a mixed valence point, but there 
$dI/dV$ curve has peaks and not steps, and their location depends on the exchange coupling between the local spin and the transport electrons\cite{JFR07}.
 
Thus, within the 3 fermion approach the low bias steps found experimentally 
must arise from higher order cotunneling processes \cite{Timm}.  
Here I adopt a  simpler approach using a phenomenological Hamiltonian with two types of electrons
 (tip and surface), with an effective spin flip assisted tunneling term \cite{Applebaum}:
\begin{equation}
{\cal H}= {\cal H}_{\rm tip} + {\cal H}_{\rm sur} + {\cal H}_{\rm S}+{\cal H}_{\rm tun}
\end{equation}
The first three terms  describe the electrons in the tip $({\cal H}_{\rm tip}=\sum_{k,\sigma} \epsilon_k  a^{\dagger}_{k\sigma}a_{k\sigma})$ and in the surface 
$({\cal H}_{\rm sur}=\sum_{p,\sigma} \epsilon_k  b^{\dagger}_{p\sigma}b_{p\sigma})$.
The  Hamiltonian of the central spin(s) is the sum of the intraatomic  ${\cal H}_{\rm 0S}(i)$ terms and 
the spin-spin couplings: :
\begin{equation}
{\cal H}_{\rm S}=\sum_i {\cal H}_{\rm 0S}(i) +\frac{1}{2}\sum_{i,j,a} j_{ab}(i,j)
\hat{S}_a(i)\cdot\hat{S}_b(j)
\end{equation}
The eigen-energies and  eigenstates of ${\cal H}_{\rm S}$ are denoted
by  $=E_M$ and $|M\rangle$.  
The tunneling terms can be written as:
\begin{equation}
{\cal H}_{ tun}=
\sum_{kk'\sigma\sigma'\alpha} T_{\alpha}(kk')
\frac{\tau^{\alpha}_{\sigma\sigma'}}{2}  \hat{S}^{\alpha}(1) 
\left(a^{\dagger}_{k\sigma} b_{k'\sigma'} + {\rm h. c.} \right)
\label{HTUN}
\end{equation}
where $\tau^{\alpha}$ and $\hat{S}^{\alpha}(1)$ are the   Pauli matrices
and  the spin operators of the spin (1) in the central region
 for $\alpha=a=x,y,z$,   and the unit matrix for $\alpha=0$.
 Because of the short-range nature of exchange interaction and tunneling processes from the tip, we assume that only one  spin $\hat{S}_{\alpha}(1)$ is
 assisting the tunneling. This term describes  the tunneling of electrons between tip and surface assisted by exchange interaction with the spin in the middle. A similar term has been used by different authors
 \cite{Nussinov03}.

Eq. (\ref{HTUN}) describes both spin assisted $(\alpha=x,y,z)$ and conventional
tunneling $\alpha=0$.  To lowest order in ${\cal H}_{ tun}$, the current has three conributions: 
(i)  a central-spin independent
$T_0^2$, 
(ii)   a crossed contributions proportional to 
 $T_0T_{a} \vec{m}_{T,S} \cdot\langle \vec{S}(1) \rangle$, and (iii) a spin flip contribution $T_a^2$ that features
the spin spectral weight  and is responsible of the steps in the $dI/dV$ curves. In the case of magnetized tip and sample the $T_0^2$ contribution depends on $\vec{m}_T\cdot\vec{m}_S$ which makes possible the SP-STS spectroscopy \cite{Wiesendanger,wortmann-PRL01}. In this work $\vec{m}_S=\vec{m}_T=0$.
The spin-flip contribution arises from
 the balance between electrons tunneling from tip to surface and back 
\cite{Applebaum}:
\begin{equation}
I=\sum_M P(M)\left( \sum_{p,\sigma}  n^T_p  \gamma_{p,M}^{T\rightarrow S}
- \sum_{k,\sigma} n^S_k   \gamma_{k,M}^{S\rightarrow T}\right)\end{equation}
where $P(M)$ is the equilibrium occupation of the $M$ state, $n^{T,S}_p$ is the occupation function of the tip and surface and $\gamma$ are the tunneling rates associated to the spin-flip assisted tunneling Hamiltonian:   
\begin{eqnarray}
\gamma_{p,M}^{T\rightarrow S}
=\sum_{p',M',a} 
|T_{a}(pp')|^2   |\langle M| \hat{S}^{a}(1)|M'\rangle|^2\times
\nonumber \\
\times \left(1-n^S_{p'}\right) 
\delta\left(\epsilon_{p'}+\epsilon_{M'}-\epsilon_{p}-\epsilon_{M}\right)
\label{gamma2c}
\end{eqnarray}
This expression gives the lifetime of a  product state with an electron in the state $p$ of the tip and the magnetic atom(s) in state $M$ due to a spin flip assisted tunneling of the electron to the surface. Importantly, this equation  relates current  
to the spin matrix elements, 
$|\langle M| S^{a}|M'\rangle|^2\equiv |S^{a}_{M,M'}|^2  $, as reported in the experiments \cite{Heinrich2,Heinrich3,Xue08}. 

If the coupling between  transport electrons and spins is rotationally invariant,
$T_{a}(k,k')=T_S$ is the same for $a=x,y,z$. The dependence of
$T_S$ on the momentum indexes can be neglected\cite{Mahan}. We take $n^T(\epsilon)=f(\epsilon)$
and $n^S(\epsilon)=f(\epsilon+eV)$, where $f$ is the Fermi function. The
sum over momenta leads to an integral over energies featuring the density of states of tip and surface,
$\rho_T(\epsilon)$ and $\rho_S(\epsilon)$ which are assumed to be flat in the neighbourhood of the Fermi energy. We arrive to 
\begin{eqnarray}
I= G_{S} \sum_{a=x,y,z}
\int_{-\infty}^ {\infty}\int_{-\infty}^ {\infty} 
{\cal S}_{aa}(\epsilon-\epsilon')F(\epsilon,\epsilon',eV)d\epsilon d\epsilon '
\label{current-spectral}
\end{eqnarray}
where $e G_{S}\equiv T_S^2 \rho_T(\epsilon_F) \rho_S(\epsilon_F)$ and 
$$
F(\epsilon,\epsilon',\omega)=\left[ f(\epsilon)\left(1-f(\epsilon'+\omega)\right) - f(\epsilon+\omega)\left(1-f(\epsilon')\right) \right]
$$
and
\begin{equation}
{\cal S}_{aa}(\omega)\equiv
 \sum_{M,M'}  
P_M |S^{a}_{M,M'}(1)|^2
\delta\left(\omega +\epsilon_{M}-\epsilon_{M'}\right)
\label{spin-spin}
\end{equation}
is the {\em spin spectral weight}.

From the formal point of view, equation (\ref{current-spectral}) is one of 
the main results  of this paper. It relates the inelastic current to the spin spectral weight of the magnetic atom  probed by the STM. It shows that
 two types of spin assisted  tunneling processes contribute to the current.
If we choose $z$ as the quantization axis,  the $a=x,y$  terms involve spin exchange between the transport electron and the magnetic atom. These are the spin flip terms.
In contrast, the $a=z$ term conserves the spin of both carrier and atom. This dichotomy is absent in the case of vibrational inelastic spectroscopy. 

One of the integrals in eq. (\ref{current-spectral}) is done using
the delta functions in   eq. (\ref{spin-spin})   and the other using
$\int_{-\infty}^ {\infty} 
\left[ f(\epsilon)\left(1-f(\epsilon+\omega)\right)\right]d\epsilon=
\frac{\omega}{1-e^{-\beta\omega}}$.
The total inelastic current is thus written as
\begin{eqnarray}
I=  \sum_{M,M',a} P_M
|S^{a}_{M,M'}(1)|^2 i(eV,\Delta_{M',M})
\label{current3}
\end{eqnarray}
with $\Delta_{M',M}=E_{M'}-E_{M}$ and
\begin{eqnarray}
 i(eV,\Delta)\equiv G_{S}
\left[\frac{eV-\Delta}{1-e^{-\beta(eV-\Delta)}}
+\frac{eV+\Delta}{1-e^{\beta(eV+\Delta)}}\right]
\end{eqnarray}
is the current associated to a single  inelastic channel with energy $\Delta$.
Equations (\ref{current-spectral}) and (\ref{current3}) are the magnetic analog of the  vibrational inelastic tunneling spectroscopy\cite{Scalapino} in which the dipole spectral weight of the molecular vibrations is replaced by the spin spectral weight of the magnetic atoms

Now the validity of  eqs. (\ref{current-spectral},\ref{current3}) and the spin assisted tunneling term (\ref{HTUN})  is verified by comparing their predictions with the experimental results.
 The case of tunneling through a single Fe atom in CuN/Cu system\cite{Heinrich3} is considered first. Following that reference, the standard single  spin Hamiltonian reads 
\begin{equation}
{\cal H}_{\rm S}= D \hat{S}_z^2 + E(\hat{S}_x^2-\hat{S}_y^2)+ g \mu_B \vec{B}\cdot\vec{S}
\label{Fe1hamil}
\end{equation}
with $S=2$ adequate for Fe$^{2+}$ and  $D=-1.55$ meV and $E=0.35$ meV\cite{Heinrich3}. 
Since $D>>E$  approximate analytical expressions
 yield the ground state energy (for $B=0$) $E_0=-4D$ and
the excitations $\Delta_{M',0}$ 
$\frac{3E^2}{D},3D-6E,3D+6E,4D$ (see fig. 2c) . 
The ground (first excited) state is made   mainly (only) with $M_z=\pm 2$ .  
The   $dI/dV$ curves obtained from eq. (\ref{current3})
and the exact solution of Hamiltonian (\ref{Fe1hamil})
 for different intensities and orientations of the applied magnetic field, evaluated for $k_BT=0.5 $K, are shown in fig.2(a,b).  The a (b) panel corresponds to field parallel to $z$ $(x)$.
At zero field the $dI/dV$ curves shows three steps corresponding to the excitations to the first, second and third excited states. The transition to the fourth state is forbidden  ( $\sum_{a=x,y,z} |\langle 0|S_{a}| 4\rangle|^2=0$). 
The prominent $0\rightarrow1$ transition comes from the spin-conserving channel $a=z$, where as the $0\rightarrow2$  and $0\rightarrow3$ transitions come from the spin flip channels  $a=y$ and $a=x$ , respectively.
Interestingly, the energy difference between these two transitions is exactly equal to $6E$. Thus, this parameter can be red from the experimental data. 
 Their evolution as a function of the intensity and orientation of the magnetic field give good account of the main observed experimental features \cite{Heinrich3}.  In particular, the conductance shows a significant magnetic anisotropy, related to that of the iron atom in this surface. 

\begin{figure}
[hbt]
\includegraphics[width=3.2in]{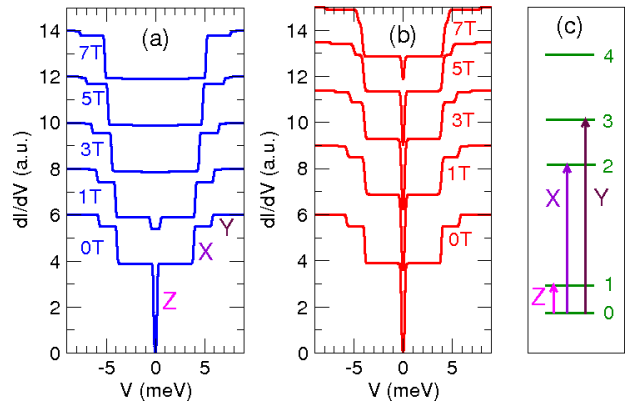}
\caption{ \label{fig2}(Color online). (a) Differential conductance for single Fe atom.
with  magnetic field along the $z$ axis.(b) The same with 
 magnetic field along the $x$ axis.  (c) Scheme of the $B=0$ energy levels of eq.(\ref{Fe1hamil}) and the  tunneling induced transitions. }
\end{figure}

The theory  also accounts for more complicated experimental situations 
where electrons tunnel through one magnetic atom which is exchanged coupled to others.
This is the case of linear chains of $N$ Mn atoms deposited on a CuN/Cu surface, $N$ going from 1 to 10. 
Mn$^{+2}$  has $S=5/2$ and  very weak magnetic anisotropy. The Mn-Mn coupling is approximated by a spin-rotational invariant first-neighbour Heisenberg coupling. The model reads: 
\begin{eqnarray}
{\cal H}_{\rm S}=\sum_i D_i \hat{S}_z^2(i) +g\mu_B \vec{B}\cdot\vec{S}(i)
 +J\sum_{i,a} \hat{S}_a(i)\cdot\hat{S}_a(i+1)
\nonumber
\end{eqnarray}
where the sum in the last term runs from $i=1$ to $N-1$.  Results
are shown up to  $N=4$ spins for which the Hilbert space has $6^4$ states. The experiments\cite{Heinrich2} can be modelled taking 
 $J\simeq 6$ meV, much larger than the single Mn  $D$.  Thus, $D$ is a weak perturbation of the Heisenberg model, whose eigenstates can be labelled with the total spin ${\cal S}$ and its third component ${\cal S}_z$. Since $j>0$  the coupling is antiferromagnetic. Thus,
 even $N$ chains have $S=0$  ground states whereas 
  odd $N$ chains have degenerate ground states with multiplicity 6 weakly split by the anisotropy term $D$. Thus, the lowest energy step occurs is related to $D$ in odd $N$ chains and to $J$ in even $N$ chains.

\begin{figure}
[hbt]
\includegraphics[width=3.2in]{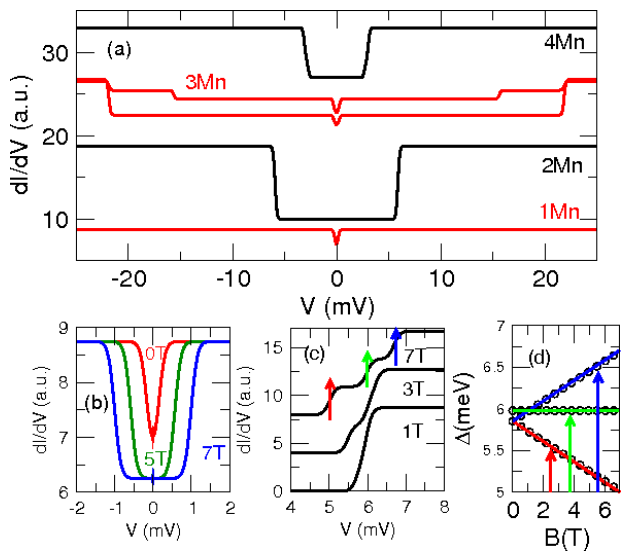}
\caption{ \label{fig3}(Color online). (a)  $dI/dV$ for $B=0$, $k_BT=0.6$K for
chains of $N$ Mn atoms, $N=1,2,3,4$. (b)  Evolution the $N=1$ low energy step  as  a function of $B$. (c)  Evolution of the first step for $N=2$ as a function of $B$.
(d) Evolution of the corresponding exciations. }
\end{figure}

The  calculations (see figure (\ref{fig3})) account  both for the difference between odd and even $N$ chains and the increase of the anisotropy related step of  odd-$N$ chains experimentally observed $dI/dV$ \cite{Heinrich2}.
The first excitation for $N=1$ and $N=3$ correspond to the transitions from the ground state doublet ${\cal S}_z=\pm 5/2$ to the first excited state with  ${\cal S}_z=\pm 3/2$
through the spin-flip channels $a=x,y$. Application of a magnetic field increases the splitting between these states and shifts the step towards higher energy, as seen in the experiment\cite{Heinrich2} and well captured by the model  (fig 3b).  

The first excited states of the dimer   have 3-fold degeneracy weakly split into a low energy doublet and a higher energy singlet  because of the single atom anisotropy. This fine structure splitting is $6.4D$.
The ground state singlet is connected to the excited state singlet via a spin-conserving transition and to the excited state doublet via spin flip transition. Figures 3c and 3d show how this fine structure evolves  as a  magnetic field is applied. The model accounts for the deviation from the Zeeman splitting  observed experimentally\cite{Heinrich2}, with $D=0.08$ meV and $J=5.89$meV.  Interestingly, the sign and magnitude of $D$ for the dimer is different than that of the monomer.

The numerical simulation  also accounts for the red shift of the first step for even-$N$ chains as $N$ increases\cite{Heinrich2}. 
This result can be rationalized using the Lieb, Shulz, Mattis theorem\cite{Lieb-Schultz-Mattis} which states that 
the energy difference of the lowest energy excitation of 
linear chains with half-integer spin and the singlet ground state (even $N$) must
is bounded by  $\frac{2\pi^2J S^2}{N}$ . Thus, the excitation energy is expected to decay as $1/N$. 

In chains with $N=3$ or  more atoms the amplitude of the spin excitations, and the height of the steps in the $dI/dV$ curves thereby,  can  vary from atom to atom. In figure 3a the $dI/dV$ curve is shown for the  spin-assisted current recorded on top of the central atom and one of the side atoms. As reported in reference (\onlinecite{Heinrich2}) the location of the steps is the same. However, the intensity is clearly different. In the spectrum recorded in the side atom two steps are seen, corresponding to transitions to the first $S=3/2$ and second $S=7/2$ excited states whereas
the spectrum recorded in the central atom only the latter is seen. Thus, the spin-assisted tunneling spectroscopy can be used to map the amplitude of the spin excitations. 

In summary,  
the experiments of  single spin inelastic tunneling spectroscopy imply  the existence of a 
 spin-assisted tunneling mechanism that  couples electrons in the tip and the surface to  the local spin (\ref{HTUN}). 
This term naturally leads to an expression for the current that involves the spin spectral weight ${\cal S}_{aa}(\omega)$ of the magnetic atom  that assist the tunnel process and to  the spin selection rules observed experimentally\cite{Heinrich2,Xue08}. 
Both the microscopic origin of the spin assisted tunneling mechanism, and its  connection to atomic scale current driven spin torque 
 deserve additional work. The spin-assisted tunneling Hamiltonian is a genuine many-body process, since  it involves four fermions, unlike ordinary tunneling. 
 It could arise from the exchange part of the electron-electron  repulsion in the Hamiltonian \cite{Applebaum} and it could be a kinetic exchange term \cite{Schrieffer-Wolff}.


I acknowledge fruitful discussions with R. Aguado, J. J. Palacios, F. Delgado, C. Untiedt and C. Hirjibehedin.  This work has been financially supported by MEC-Spain (Grant
MAT07  ) and  by Consolider CSD2007-0010.

{\em Note added:}. During the  completion of this work, a preprint with related results has been posted \cite{Balatsky08}.



\newpage







\begin{references}

 
\bibitem{Josephson} B. D. Josephson, Rev. Mod. Phys. {\bf 36}, 217 (1964)

\bibitem{TMR}  
M. Julli\`ere, Phs. Lett. {\bf 54}A, 225 (1975).
J. S. Moodera, L. R. Kinder, T. M. Wong, R. Meservey, Phys. Rev. Lett.{\bf 74}, 3273 (1995).
\bibitem{Jaklevic66}  R. C. Jaklevic and J. Lambe, Phys. Rev. Lett. {\bf 17}, 1139  
(1966)

\bibitem{Tedrow} P. M. Tedrow, R. Meservey, Phys. Rev. B{ \bf 7}, 318 (1973)


\bibitem{STM}  G. Binning and H. Rohrer, Rev. Mod. Phys.{\bf 59}, 615 (1987)



\bibitem{Scalapino} D. J. Scalapino, S. M. Marcus, 
Phys. Rev. Lett. {\bf 18}, 459    (1967 )

 
\bibitem{Ho} B. C. Stipe,M. A. Rezaei, W. Ho , Science {\bf 280}, 1732 (1998)

\bibitem{BGR} G. Binnig, N. Garcia, H. Rohrer, Phys. Rev. B{\bf 32}, 1336 (1985)


\bibitem{Heinrich1} A. J. Heinrich, J. A. Gupta, C. P. Lutz, D. M. Eigler, Science {\bf 306}, 466 (2004)

\bibitem{Heinrich2} C. F. Hirjibehedin, C. P. Lutz, A: J. Heinrich, Science {\bf 312}, 1021 (2006)

 \bibitem{Heinrich3} C. Hirjibehedin, C-Y Lin. A.F. Otte, M. ternes, C. P. Lutz, B. A. Jones, A. J. Heinrich, Science {\bf317}, 1199 (2007) 


\bibitem{Xue08} Xi. Chen {\em et al.}, Phys. Rev. Lett. {\bf 101}, 197208 (2008)

\bibitem{Wiesendanger} F. Meier, L. Zhou, J. Wiebe, R. Wiesendanger, 
Science {\bf 320}, 82 (2008)

\bibitem{persson} M. Persson, arXiv:0811.2511 



\bibitem{JFR07} J. Fern\'andez-Rossier and R. Aguado, Phys. Rev. Lett. {\bf 98}, 106805 (2007)
 

\bibitem{Timm} F. Elste, C. Timm, Phys. Rev. B{\bf 75}, 195341 (2007)

\bibitem{Applebaum} J. A. Applebaum, Phys. Rev. {\bf 154}, 633 (1967)

\bibitem{Nussinov03} Z. Nussinov, M. F. Crommie, A. V. Balatsky, Phys. Rev. B{\bf 68},
085402, (2003). G. H. Kim, T. S. Kim, Phys. Rev. Lett. {\bf 92}, 137203 (2004)



\bibitem{wortmann-PRL01}  D. Wortmann, S. Heinze, Ph. Kurz, G. Bihlmayer, S. Bl\~ugel, 
Phys. Rev. Lett. {\bf 86}, 4132 (2001)

\bibitem{Mahan} G. Mahan, {\em Many particle Physics}, Plenum, second edition (1992), chapter 9.3











\bibitem{Lieb-Schultz-Mattis} E. Lieb, T.D. Schultz and D. C. Mattis, Ann. Phys. {\bf 16}, 407 (1961). A. Auerbach, {\em Interacting electrons and quantum magnetism}, Springer, 1994. 

\bibitem{Schrieffer-Wolff} 
J. R. Schrieffer and P. A. Wolf,  Phys. Rev. {\bf 149}, 491 (1966)

\bibitem{Balatsky08} J. Fransson, O. Eriksson, A. V. Balatsky, arXiv:0812.4956





\end{references}
\end{document}